\documentclass[a4paper]{jpconf}
\usepackage{graphicx}
\usepackage{amsfonts}
\usepackage{tikz}
\DeclareRobustCommand\full  {\tikz[baseline=-0.6ex]\draw[thick] (0,0)--(0.5,0);}
\DeclareRobustCommand\dashed{\tikz[baseline=-0.6ex]\draw[thick,dashed] (0,0)--(0.54,0);}

\usepackage{hyperref}
\usepackage[percent]{overpic}

\begin{document}
\title{Prediction of wall-bounded turbulence from wall quantities using convolutional neural networks
}

\author{Luca Guastoni$^{1,2}$, Miguel P. Encinar$^3$, Philipp Schlatter$^{1,2}$, Hossein Azizpour$^{4,2}$ and Ricardo Vinuesa$^{1,2}$}

\address{$^1$ Linn\'e FLOW Centre, KTH Mechanics, Stockholm, Sweden}
\address{$^2$ Swedish e-Science Research Centre (SeRC), Stockholm, Sweden}
\address{$^3$ School of Aeronautics, Universidad Polit\'ecnica de Madrid, Madrid, Spain}
\address{$^4$ Division of Robotics, Perception, and Learning, School of EECS, KTH Royal Institute of Technology, Stockholm,
Sweden}

\ead{guastoni@mech.kth.se}

\begin{abstract}
A fully-convolutional neural-network model is used to predict the streamwise velocity fields at several wall-normal locations by taking as input the streamwise and spanwise wall-shear-stress planes in a turbulent open channel flow. The training data are generated by performing a  direct numerical simulation (DNS) at a friction Reynolds number of $Re_{\tau}=180$. Various networks are trained for predictions at three inner-scaled locations ($y^+ = 15,~30,~50$) and for different time steps between input samples $\Delta t^{+}_{s}$. The inherent non-linearity of the neural-network model enables a better prediction capability than linear methods, with a lower error in both the instantaneous flow fields and turbulent statistics. Using a dataset with higher $\Delta t^+_{s}$ improves the generalization at all the considered wall-normal locations, as long as the network capacity is sufficient to generalize over the dataset. The use of a multiple-output network, with parallel dedicated branches for two wall-normal locations, does not provide any improvement over two separated single-output networks, other than a moderate saving in training time. Training time can be effectively reduced, by a factor of 4, via a \textit{transfer learning} method that initializes the network parameters using the optimized parameters of a previously-trained network. 
\end{abstract}

\section{Introduction}
After becoming the state-of-the-art algorithm for cognitive tasks such as visual recognition and natural language understanding, machine-learning techniques are gaining acceptance in many disciplines of natural science and engineering, where the knowledge of the underlying physics represents both an additional challenge and opportunity. In particular, the interest in applications to fluid dynamics has been increasing in the recent years, as summarized in Ref.~\cite{brunton2019}. Modelling tools like neural networks have been used, among others, to develop turbulence-closure models for Reynolds-averaged Navier--Stokes (RANS) simulations~\cite{ling_et_al}, flow control~\cite{Rabault} and machine-aided turbulence theory~\cite{jimenez_ml}.  
The phenomena that characterize turbulence in the near-wall region have also been modeled with the aid of neural networks: in our previous studies~\cite{srinivasan, guastoni_tsfp} multilayer perceptrons (MLPs)~\cite{rumelhart1985learning} and long short-term memory (LSTM) networks~\cite{hochreiter1997long} have been used to predict the temporal evolution for a low-dimensional representation of the near-wall cycle of turbulence~\cite{moehlis_et_al}. 

Our objective in this work is to predict instantaneous turbulent flow fields based on wall measurements. To this end, several methods have been proposed in the past. Despite the presence of both linear and non-linear interactions in turbulent flows \cite{agostini_leschziner_2016,baars_et_al_2017,dogan_et_al_2019}, most works provide a linear approximation, which allows to take advantage of the extensive work on linear systems analysis in the literature~\cite{dullerud}. For instance, Illingworth {\it et al.}~\cite{illingworth_et_al_2018} employed a linear dynamical-system approach based on the resolvent-analysis framework~\cite{mckeon_sharma_2010} to predict the velocity field on a horizontal plane in the logarithmic region of a turbulent channel flow, based on the velocity field from another horizontal plane in the same region. On the other hand, Suzuki and Hasegawa~\cite{suzuki_hasegawa_2017} and Encinar {\it et al.}~\cite{encinar}, employed linear stochastic estimation (LSE) to reconstruct different horizontal planes of the flow in a turbulent channel based on wall measurements such as the pressure, and the two components of the wall-shear stress. Sasaki {\it et al.}~\cite{sasaki_vinuesa_cavalieri_schlatter_henningson_2019} have recently assessed flow-prediction methods based on single- and multiple-input linear transfer functions, which can then be used as convolution kernels to predict the fluctuations in a spatially-developing turbulent boundary layer. In particular, they performed predictions of the near-wall flow based on horizontal velocity fields in the outer region, and they also predicted the flow based on wall measurements.  Note that Sasaki {\it et al.}~\cite{sasaki_vinuesa_cavalieri_schlatter_henningson_2019} also documented significant improvements in the predictions when using non-linear transfer functions to relate the input and the output.

In this study a variant of artificial neural networks (ANN), namely fully-convolutional neural networks, are used to predict the streamwise velocity field in an open-channel flow at one (\textit{single-output}) or more (\textit{multiple-output}) distances from the wall, using the wall-shear stress in streamwise and spanwise directions as input. ANNs have already been applied to a related problem, in one of the earliest applications of neural networks to fluid dynamics~\cite{milano_koumoutsakos}. In that work, a fully-connected NN was trained to improve a second-order-accurate model of the near-wall flow, by introducing a higher-order term.
More recently, Guemes {\it et al.}~\cite{guemes} considered ANNs to predict the temporal coefficient of the most energetic modes of the flow, which were obtained via proper orthogonal decomposition (POD)~\cite{Lumley}. In that case, the ANN outperformed the so-called extended POD~\cite{boree} thanks to its capability of reproducing the non-linear features of the flow. 
 
This manuscript is organized as follows: in Section~\ref{s:meth} we describe the simulation that has been used to generate the training and test datasets, as well as the considered neural-network architectures. 
In that section we also discuss the application of transfer learning. 
In Section~\ref{s:results} we assess the performance of the various trained models when predicting both instantaneous fields and turbulence statistics, in comparison with LSE. 
The results of the multiple-output architecture and with transfer learning are also compared with the ones from previously-trained single-output networks. Finally, in Section~\ref{s:concl} we summarize the obtained results and we outline current challenges and future developments.

\section{Methodology}\label{s:meth}

\subsection{Dataset description}\label{ss:dataset}

All the ANN variants in this study have been trained using the data generated from the direct numerical simulation (DNS) of a turbulent open channel flow. Periodic boundary conditions are imposed in the \textit{x-} and \textit{z-}directions (which are the streamwise and spanwise coordinates, respectively), and a no-slip condition is applied at the lower boundary ($y=0$, where $y$ is the wall-normal coordinate). Differently from a standard channel flow simulation, a symmetry condition is imposed at the upper boundary. 
In standard channel flows, the effect of the upper wall extends way beyond the middle plane of the flow~\cite{lozano_2012}. On the other hand, in open channel flows the upper wall is not present, making the simulation more suitable to investigate to which extent the neural networks are able to learn the dynamics of near-wall turbulence, since the interaction of the large scales with both walls is not present. 

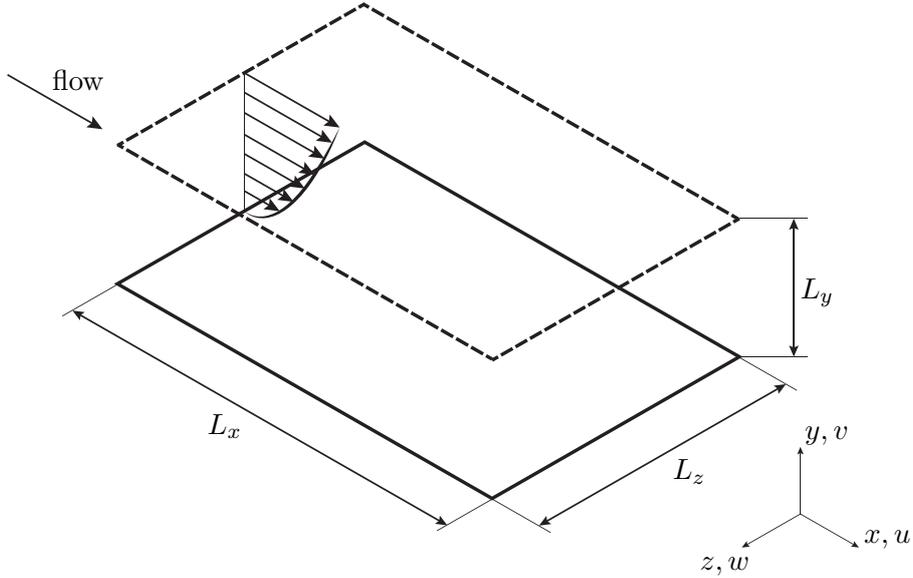
\begin{figure}
\begin{center}
\begin{overpic}[width=0.75\textwidth]{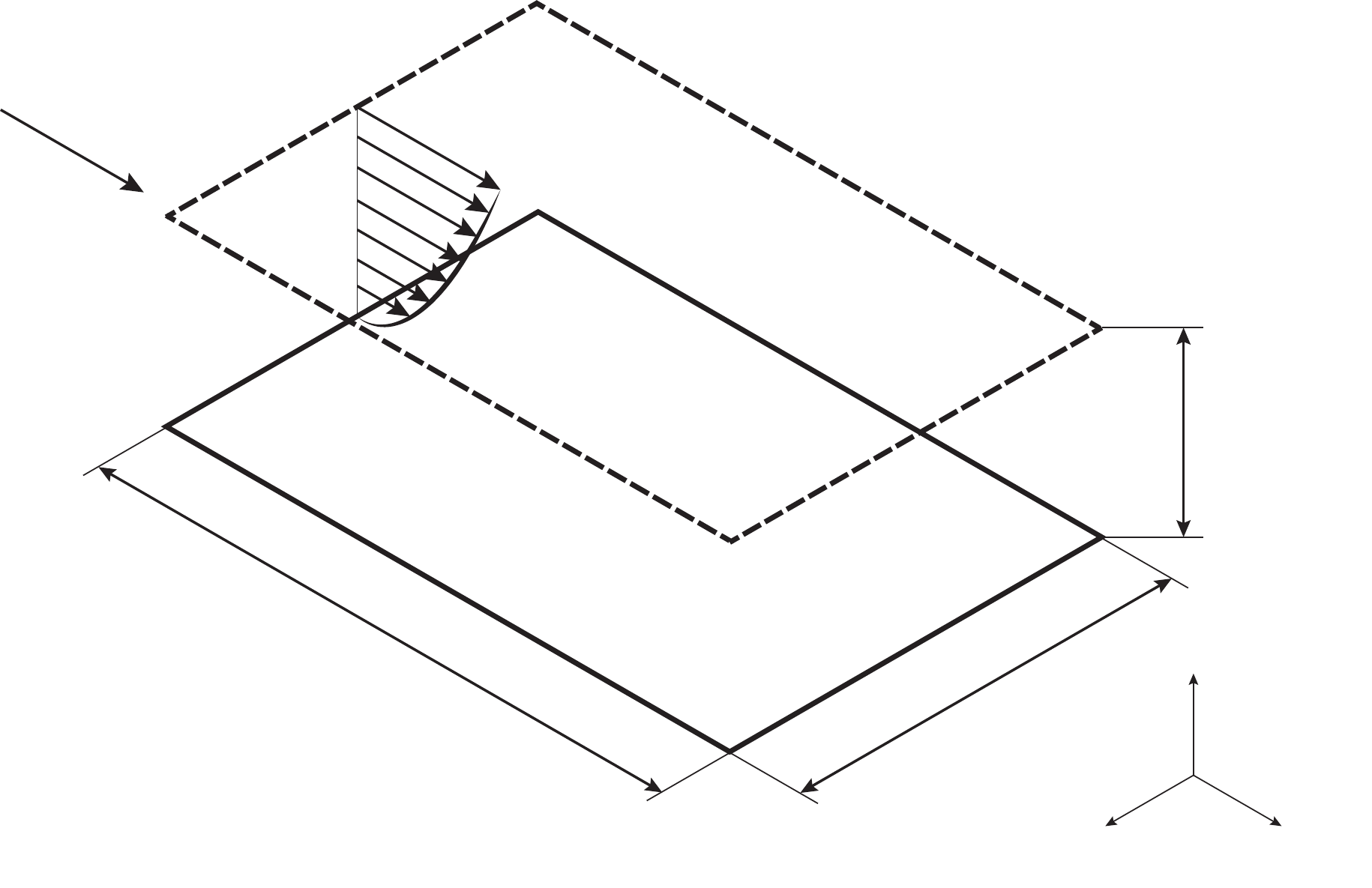}
 \put (5,55) {flow}
 \put (22,17) {$L_x$}
 \put (73,12) {$L_z$}
 \put (87,32) {$L_y$}
 \put (87.5,16.5) {$y,v$}
 \put (94,5) {$x,u$}
 \put (76,2) {$z,w$}
\end{overpic}
\end{center}
\caption{\label{channel}Computational domain and frame of reference for the DNS of the turbulent open channel considered in this study.}
\end{figure}

The simulation is performed using the spectral code SIMSON~\cite{chevalier}. The flow field is represented with $N_y=129$ Chebyshev modes in the wall-normal direction and with $N_x=128$ and $N_z=128$ Fourier modes in the streamwise and spanwise directions, respectively. The instantaneous fields are obtained at constant time intervals from the time-advancing scheme, which is a second-order Crank--Nicolson scheme for the linear terms and a third-order Runge--Kutta method for the non-linear terms. Dealiasing using a standard 3/2-rule is employed in the wall-parallel Fourier directions. The simulation is performed at constant mass flow rate at a friction Reynolds number (based on the channel height $h$ and the friction velocity $u_{\tau}$) of $Re_{\tau} = 180$, in a domain $\Omega = L_x \times L_y \times L_z = 2\pi h \times h \times \pi h$, as shown in Figure~\ref{channel}. 
The streamwise velocity fields to be used as ground truth for training and testing are sampled at different inner-scaled coordinates: $y^{+}=15,~30$ and $50$. Note that `+' denotes viscous scaling, {\it i.e.} in terms of the friction velocity $u_{\tau}$ or the viscous length $\ell^{*}=\nu / u_{\tau}$ (where $\nu$ is the fluid kinematic viscosity). A dataset is defined as a collection of samples, each consisting in the wall-shear-stress fields and the corresponding streamwise velocity field. The sampling time in the simulation is $\Delta t^+_{s} = 0.56$, and its impact on the results will be thoroughly tested. This parameter can significantly affect the generalization of the trained network: the larger the time interval, the less correlated the samples will be. The training of the neural networks is performed using a stochastic-gradient-descent algorithm and a lower correlation makes the training more difficult, meaning that the number of times that the algorithm has to iterate over the entire dataset (also called \textit{epochs}) will be higher, to achieve the same value of the loss function. At the same time, the risk of overfitting to the training data is reduced. 

\subsection{Convolutional neural network}\label{ss:nn}
Multilayer Perceptrons (MLP), also known as fully-connected networks, became popular after the invention of the backpropagation algorithm~\cite{rumelhart1985learning}. Even with backpropagation, it can still be challenging to optimize an MLP with multiple hidden layers, mainly due to the large number of learnable parameters (also called \textit{weights}) and the gradient vanishing through multiple layers. Instead, there is another variant of ANN, which adopts convolution operations in each layer of the network, which is more commonly used for practical applications with high input dimensions. This variant is referred to as \textit{convolutional neural network} (CNN)~\cite{Lecun98gradient-basedlearning}, and several comprehensive descriptions of CNNs are available in the literature~\cite{Goodfellow:2016:DL:3086952}. Here, we limit ourselves to a brief introduction of the main concepts and terminology. We consider convolution operations in two dimensions, defined as:
\begin{equation}
    F_{i,j} = \sum_m \sum_n I_{i-m,j-n}K_{m,n}
\end{equation}
where $\mathbf{I} \in \mathbb{R}^{d_1 \times d_2}$ is the input, $\mathbf{K} \in \mathbb{R}^{k_1 \times k_2}$ is called \textit{kernel} (or \textit{filter}) and contains the learnable parameters of the layer. The transformed output $\mathbf{F}$ is called \textit{feature map}. Multiple feature maps can be stacked and sequentially fed into another convolutional layer as input. This allows the next layer to combine the features individually identified in each feature map, enabling the prediction of larger and more complex features for progressively deeper convolutional networks. Since $k_i \ll d_i\ \forall i$, the use of kernels greatly reduces the number of parameters that need to be learned during training (in comparison with fully-connected networks).
The input region from which a single point of the output can draw information is called \textit{receptive field} and it can be computed based on the network architecture~\cite{dumoulin2016guide}.  

Since the simulations consider periodic boundary conditions in the streamwise and spanwise directions, a suitable boundary treatment is required for the predictions. 
Convolutional networks act only locally by means of their kernels, hence we modified the input of the network to address the periodicity as follows: all the input planes are padded in the periodic directions, which means that, for each direction, they are extended on both ends, using the values from the other side of the field through the periodic boundary.  In this study the padding is computed to take into account the size difference between input and output, due to the application of the convolution operations. For the network architecture in this study, 14 points are added to each $128\times128$ flow field, both in the $x$ and $z$ dimensions, obtaining a image that is around $20\%$ bigger. Note however that in our case, this is not sufficient to have a periodic output: to obtain a correct prediction of the boundary conditions, the network has to receive the same input on either side of the domain. The required padding has to be computed using the size of the receptive field of the network, thus involving a higher number of points for padding the input with respect to what is considered in this work. The assessment of this approach, allowing to obtain periodic predictions, is left for future work.   

The ANN architecture used for this study is a fully-convolutional neural network (FCN): it is composed of stacked convolutional layers and their respective non-linear activation functions. FCNs are commonly used in applications where the input and output domains have structural similarities. \textit{Image segmentation}~\cite{long2015fully} is one such case where the output has the same spatial dimensions as the input, like our case in which a two-dimensional flow field is predicted.
The inputs of the network are the wall-shear-stress components in the streamwise and spanwise directions. The output is the instantaneous streamwise velocity, denoted as $u$, at one or more given distances from the wall, at the same time as the wall-shear-stress fields. The network is trained to minimize the loss function:
\begin{equation}
\mathcal{L}(u_\mathrm{FCN};u_\mathrm{DNS})=\frac{1}{N_x N_z} \sum_{j=1}^{N_x} \sum_{i=1}^{N_z} \left | u_\mathrm{FCN}(i,j) - u_\mathrm{DNS}(i,j)\right |^{2},
\end{equation}
that is the mean squared error (MSE) between the instantaneous prediction $u_\mathrm{FCN}$ and the actual flow field $u_\mathrm{DNS}$, as computed by the DNS. 

While convolutional neural networks can be used to work on time sequences~\cite{oord2016wavenet}, our choice of the inputs and outputs allows the model to learn only spatial correlations between the different points in the wall-shear-stress fields to predict the velocity field. 
We have recently shown~\cite{guastoni_tsfp} that using a network that can learn temporal dynamics, such as a network consisting of LSTM units, can drastically improve the prediction performance. On the other hand, the use of a model without this capability is more flexible during the training stage and when inference is performed (\textit{i.e.}\ when the predictions are computed from the wall-shear-stress fields in the test dataset).
LSTM networks, as well as other models working with temporal sequences, usually assume a constant time separation between input samples, which might not be the case if we have a numerical simulation with adaptive time step. 
Unevenly spaced data require specific neural-network architecture variants, such as the recurrent neural network proposed by Neil {\it et al.}~\cite{Neil:2016:PLA:3157382.3157532}. 
During the inference stage, networks that learn temporal dynamics would typically require a sequence as input to perform the prediction. On the other hand, in our setup the FCN predicts using a single wall-shear-stress plane.

The ANNs have been trained using 25,200 instantaneous fields, split into training and validation sets, with a ratio of 4:1. We employ the Adam~\cite{kingmaba} optimization algorithm, using the parameter values suggested in the original paper and with a scheduled exponential learning-rate decay.
\begin{figure}
\begin{center}
\includegraphics[width=0.8\textwidth]{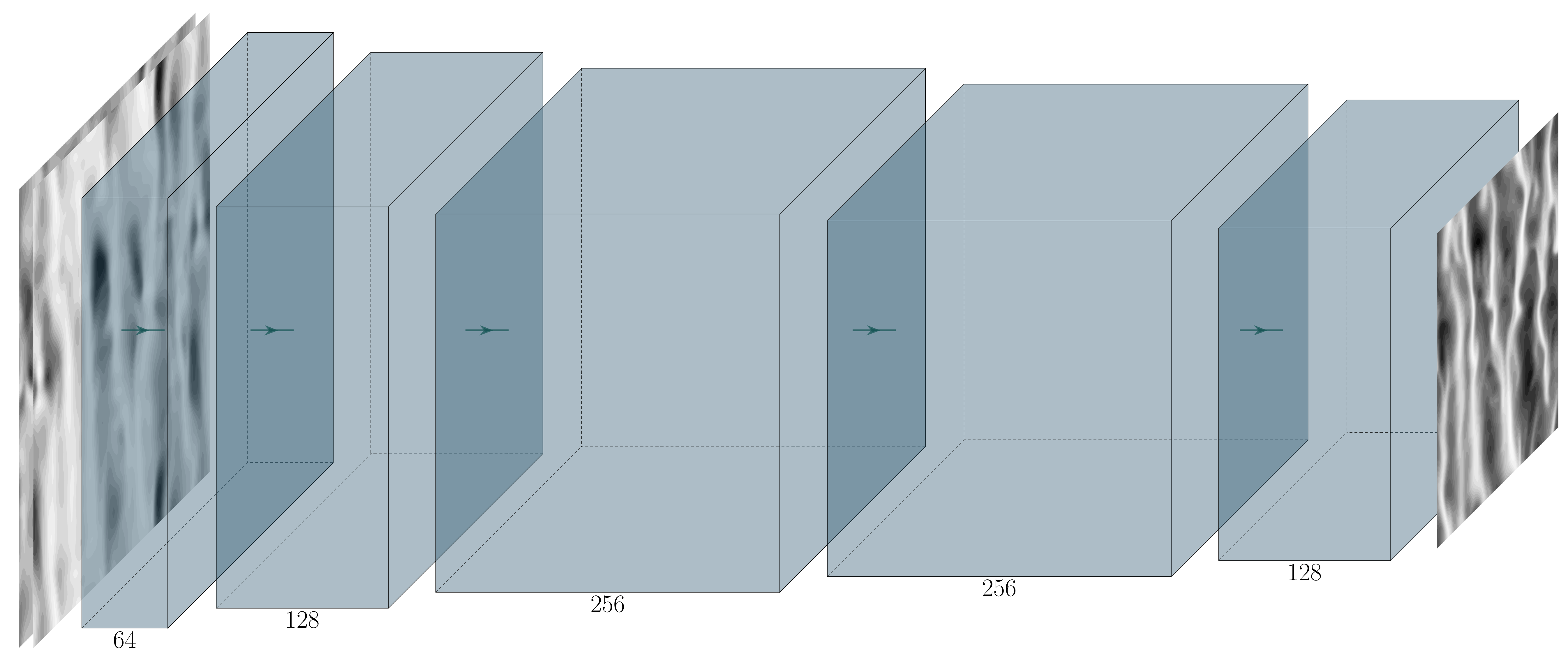}
\end{center}
\caption{\label{fig:net} Schematic representation of the considered FCN architecture. The input fields are on the left and the output on the right. The numbers indicate the number of kernels applied to each of the layers. The kernels (not represented in the Figure) have size $(5\times 5)$ in the first convolutional layer, and $(3\times 3)$ in the subsequent layers}
\end{figure}
The neural-network architecture chosen for this study is shown in Figure~\ref{fig:net}, where the number of kernels is indicated below each convolutional layer. A batch normalization~\cite{ioffe} is performed after each convolutional layer (except for the last one) and then a rectified linear unit (ReLU)~\cite{nair2010rectified} activation function is applied to the feature maps. 
Unless differently noted, only one plane at a time is predicted, which means that a different neural network has to be trained for every target wall-normal location.  

A variation of the described architecture, allowing to predict two different wall-normal locations at the same time, is also tested. 
The network consists of a common part, shared between the two outputs and two dedicated branches. The optimization algorithm updates the weights in the common part using a weighted sum of the error gradients computed for each output. In our case we considered an equal contribution from both outputs. The weights of each branch are updated only with the error associated to the respective output. This architecture is justified by the way convolutional neural networks predict the output features, as described at the beginning of this subsection. In the multiple-output architecture, the prediction of the plane farthest away from the wall may benefit from error-gradient information coming from a plane closer to the wall. 
Given the same input, we assume that the smaller features that are extracted from the first layers are similar for the different wall-normal locations, hence they can be computed only once. The two branches of the network proceed to combine the smaller features into larger ones, which are different and characteristic for each wall-normal location.  

A similar approach is used also to reduce the training time of the single-output network. Transfer learning concerns the techniques that transfer knowledge from optimized models on some related tasks to a model for a new task. In our case we used \textit{fine tuning} as a transfer-learning technique for neural networks, initializing the network for the new task with a set of weights taken from a previously-trained network with the same architecture, but a different target height. 
The initialized network is trained on a new dataset, however the error is backpropagated only in the very last layers, with a smaller learning rate than the original training. This way it is possible to optimize the composition of the smaller features into larger ones for the new dataset. Reducing the number of weights that are learnt, a much faster training is achievable, at the cost of a reduced flexibility of the model.
Transfer learning has been successfully used for many different applications~\cite{azizpour2015factors}, also for purposes that were quite different from the original ones~\cite{zhang2018unreasonable}.
In our study, the velocity fields come from the same DNS database, hence transfer learning can be effectively used to speed up the training. We initialized the network using the weights computed to predict the velocity field at the location of the near-wall fluctuation peak, {\it i.e.}\  $y^{+}=15$, and we updated the weights of the last three convolutional layers. 

\subsection{Linear stochastic estimation (LSE)}\label{ss:lse}
The trained FCNs are able to provide a nonlinear prediction of the flow field, differently from most previously implemented methods, which are based on a linear flow reconstruction. 
To quantify the improvement over linear methods, the FCN predictions are compared with a linear stochastic estimation (LSE)~\cite{adrian88}. This method was recently used to estimate the velocity field on a horizontal plane in a turbulent channel flow~\cite{suzuki_hasegawa_2017} and the second invariant of the velocity gradient~\cite{encinar}.

Stochastic estimation consists in the approximation of some unknown variables $\bi{u}$, from the knowledge of a set of observables $\bi{E}$, and their joint probability density function, $P(\bi{u}, \bi{E})$. If the probability distribution is fully known, the best estimation of $\bi{u}$ in a mean square sense is the mean of $u$ conditioned to the observed state of $E$ $\langle \bi{u} | \bi{E} \rangle$. When the full joint probability density function is not available, the approximation can be expanded in Taylor series, and is called LSE when truncated at the linear term:
\begin{equation}\label{eq:lse}
    \bi{u} \approx \mathbf{L} \bi{E},
\end{equation}
where $\mathbf{L}$ is the operator corresponding to the linear term of the Taylor expansion.
Because LSE uses all the available information to estimate the unknowns, is formally equivalent to a linear regression. Thus, $\mathbf{L}$ can be computed by solving the least-square system:
\begin{equation}\label{eq:lse2}
     \langle \bi{E}^\mathrm{T} \bi{E} \rangle\mathbf{L} = \langle \bi{u}^\mathrm{T} \bi{E} \rangle,
\end{equation}
where the left-hand-side factor corresponds to the correlation tensor of the observables and the right-hand-side factor to the cross-correlation tensor of the unknowns with the observables.
For a complete comparison, we computed the linear operator $\mathbf{L}$ using different time steps between samples, to mirror the different training sets for the neural networks.

\section{Results}\label{s:results}
We assess the performance of the trained models comparing the predicted velocity fields with the reference ones, obtained from the DNS. It is important that no spurious time and spatial correlations are present between the training and test datasets, to avoid a potential unfair advantage to the neural networks. The most convenient approach for us to enforce this was to take the samples in the testing set from a separate simulation, initialized with a random seed, different from that of the training-data simulation.
We also average over the fields obtained from the neural-network model to evaluate the statistics of the predicted flow, namely the mean velocity in the streamwise direction $\langle u \rangle^+$ and the fluctuation intensity $u_\mathrm{RMS}^+$ (where RMS denotes root-mean-squared). The statistical errors $E_{\langle u \rangle^+}$ and $E_{u_\mathrm{RMS}^+}$ are computed as follows: 
\begin{equation}
    E_{\langle u \rangle^+} = \frac{\left| \langle u_\mathrm{FCN} \rangle^+ - \langle u_\mathrm{DNS} \rangle^+\right|}{\langle u_\mathrm{DNS} \rangle^+}, \quad\quad E_{u_\mathrm{RMS}^+} = \frac{\left| u_\mathrm{RMS,NN}^+ - u_\mathrm{RMS,DNS}^+ \right|}{u_\mathrm{RMS,DNS}^+},
\end{equation}
where as above the subscripts DNS and NN refer to the reference and predicted profiles, respectively.
\begin{figure}
\begin{center}
\includegraphics[width=.328\textwidth]{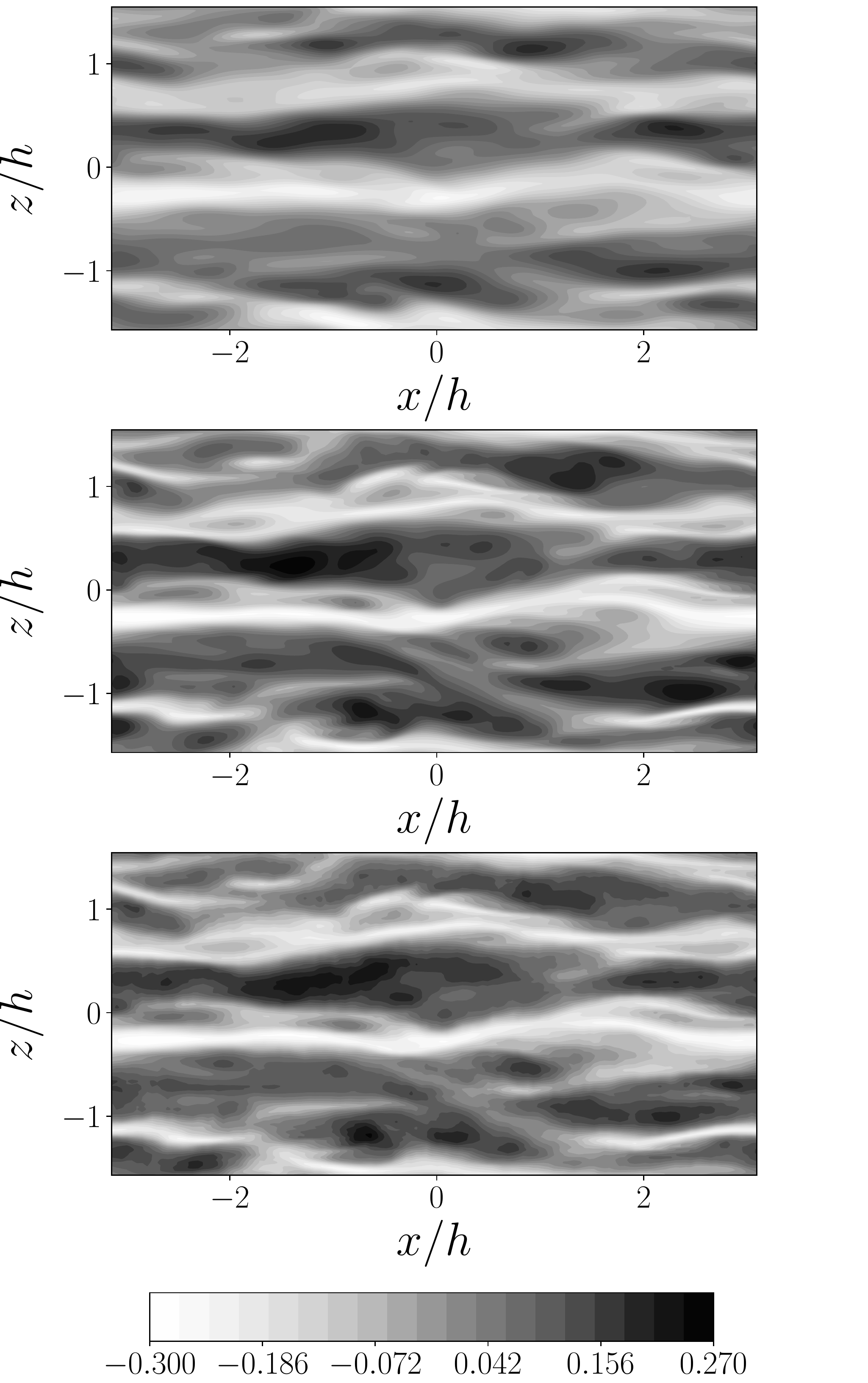}
\includegraphics[width=.328\textwidth]{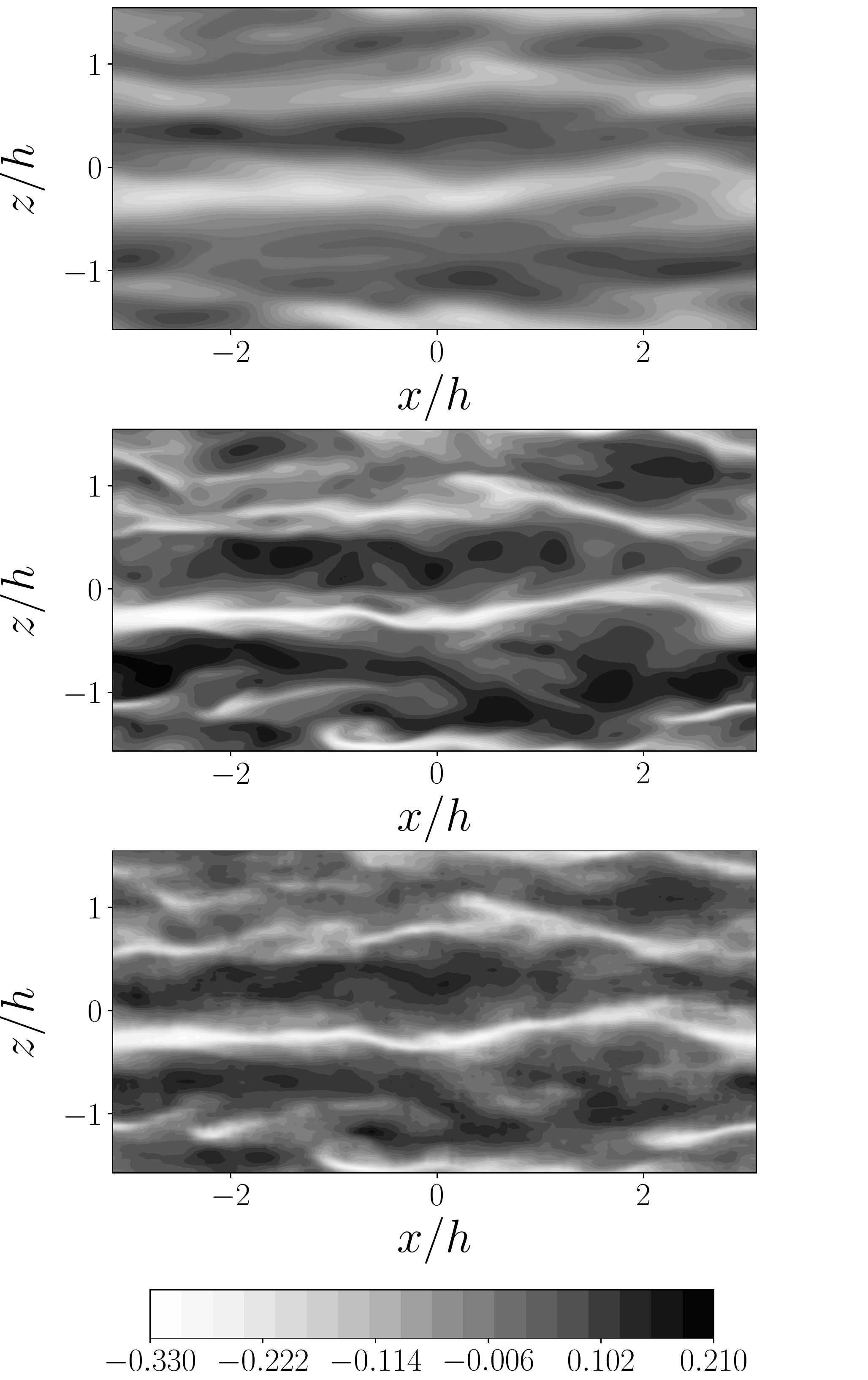}
\includegraphics[width=.328\textwidth]{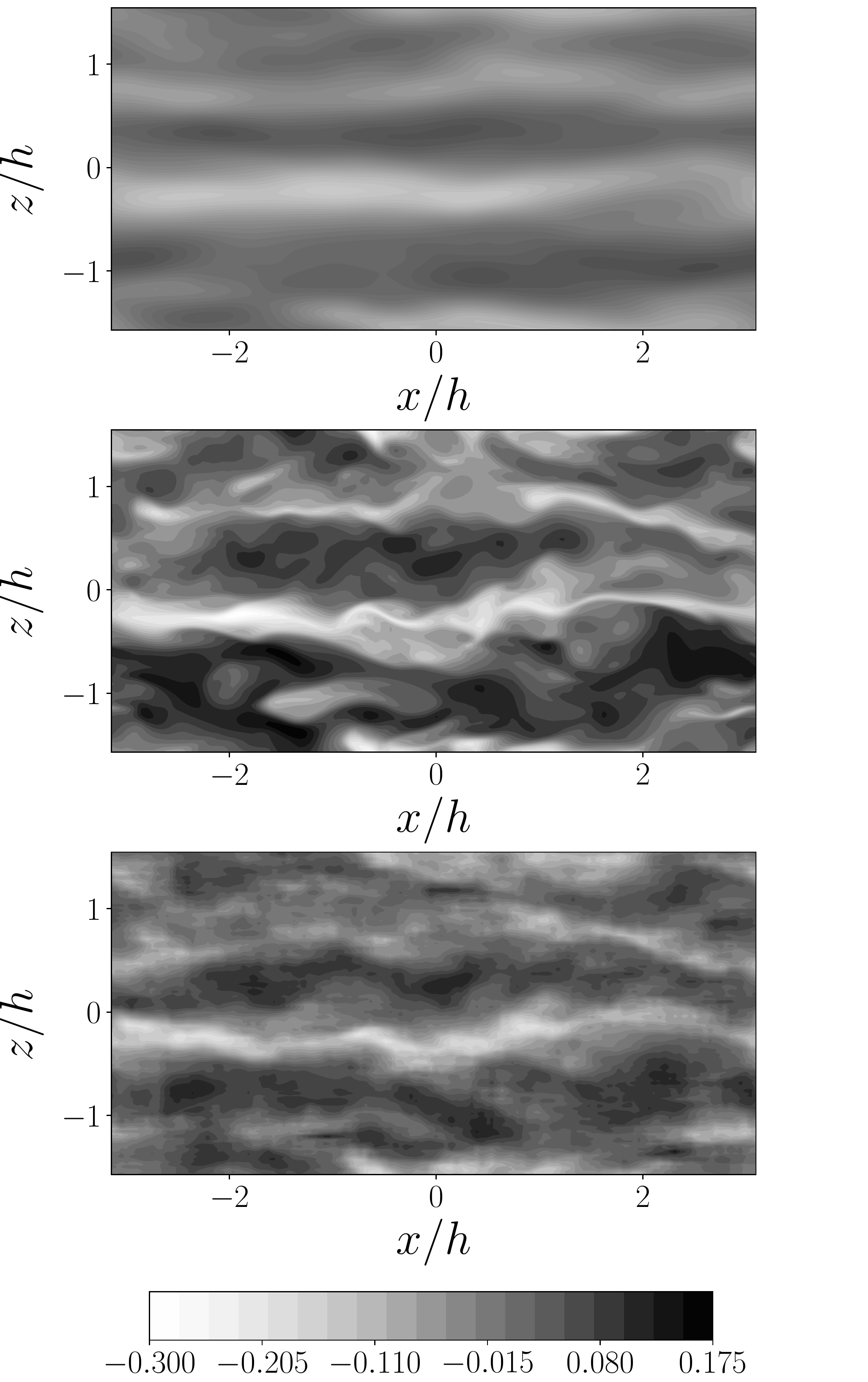}
\end{center}
\caption{\label{fig:field_comp} Comparison of the streamwise velocity-fluctuation flow fields from the LSE (top), the reference DNS (center) and the convolutional neural network (bottom), at $y^+=15$ (left), $y^+=30$ (center) and $y^+=50$ (right). The dataset with $\Delta t^+_{s} = 15.25$ was used to train the convolutional neural networks and to compute the linear operator for the LSE.} 

\end{figure}
The predictions are analyzed first from a qualitative point of view and then from a statistical one. In the latter case, since the neural network is optimized using a stochastic algorithm, the statistics at each $y^+$, for every $\Delta t^+_{s}$, are averaged over 5 models, each of them trained for 100 epochs, starting from different weight initializations taken from the same distribution. 
In Figure~\ref{fig:field_comp}, examples of the velocity fluctuations obtained from predicted fields at different distances from the wall are shown. The best-performing ANN at each height is compared with the LSE, which provides a linear reconstruction. At $y^+=15$, both the LSE and the neural network are able to provide accurate results, where turbulent structures like the high- and low-speed streaks that characterize the flow at this distance from the wall are clearly identifiable (note however that the LSE reconstruction exhibits a small attenuation in the predicted fluctuations, as well as smoother near-wall streaks, compared with the reference data). At $y^+=30$, the neural network maintains a comparable level of accuracy, while in the LSE reconstruction some of the higher frequency fluctuations are not captured. The performance degrades when moving away from the wall, and at $y^+=50$ only the neural network is able to provide a prediction resembling the structure of the original data, in particular in the regions of the plane where the velocity gradient is stronger. Note that the two algorithms differ also in the way that the loss of accuracy affects the predictions: the performance degradation of the LSE results in flow fields that are smoother than the reference data, with reduced velocity fluctuations. On the other hand, the predictions of the neural network have more irregular velocity contours, thus providing a somewhat noisier representation of the flow fields.

Investigating the local performance of the neural network, we found that at all heights the predicted velocity is overestimated where the original velocity is lower, and underestimated when it is larger. This behavior can be directly related to the optimization based on the minimization of the mean-squared error, which tends to smooth out peak values.    

\begin{figure}
\begin{center}
\includegraphics[width=.47\textwidth]{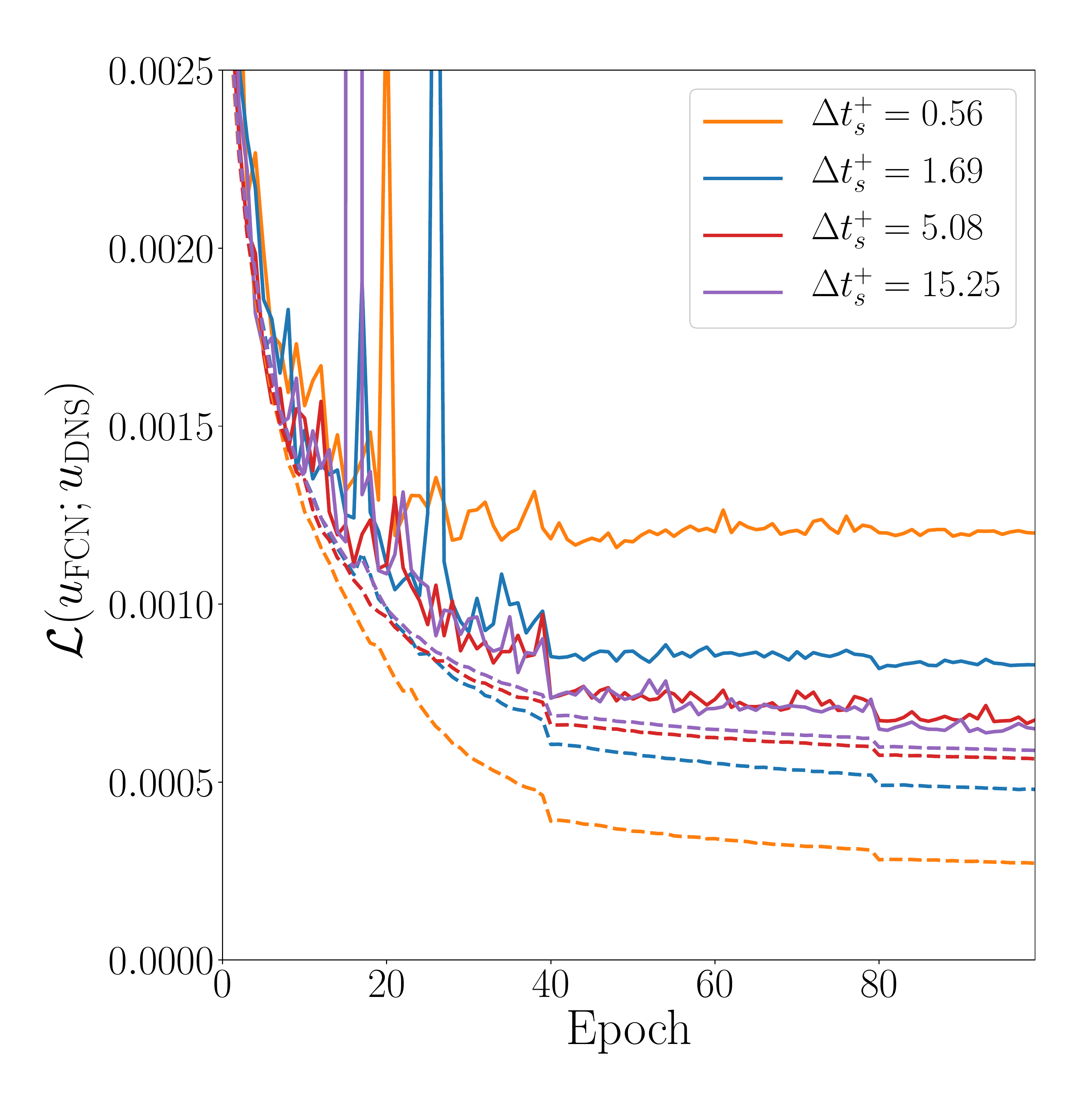}
\includegraphics[width=.47\textwidth]{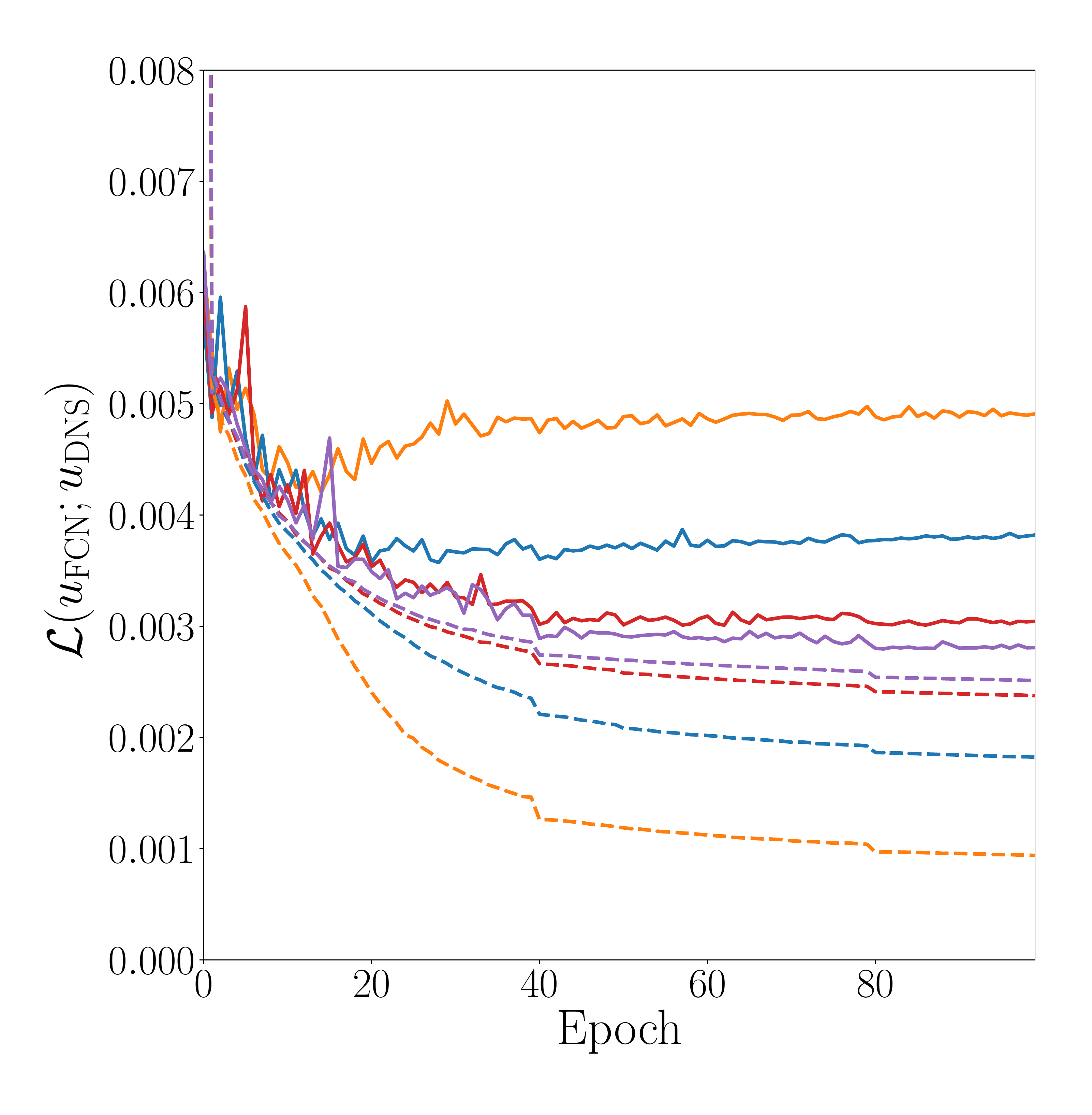}
\end{center}
\caption{\label{fig:tstep}Training (\dashed) and validation (\full) loss in the FCN prediction at $y^+=15$ (left) and $y^+=50$ (right), using different time steps between samples.}
\end{figure}
\begin{figure}
\begin{center}
\includegraphics[width=.328\textwidth]{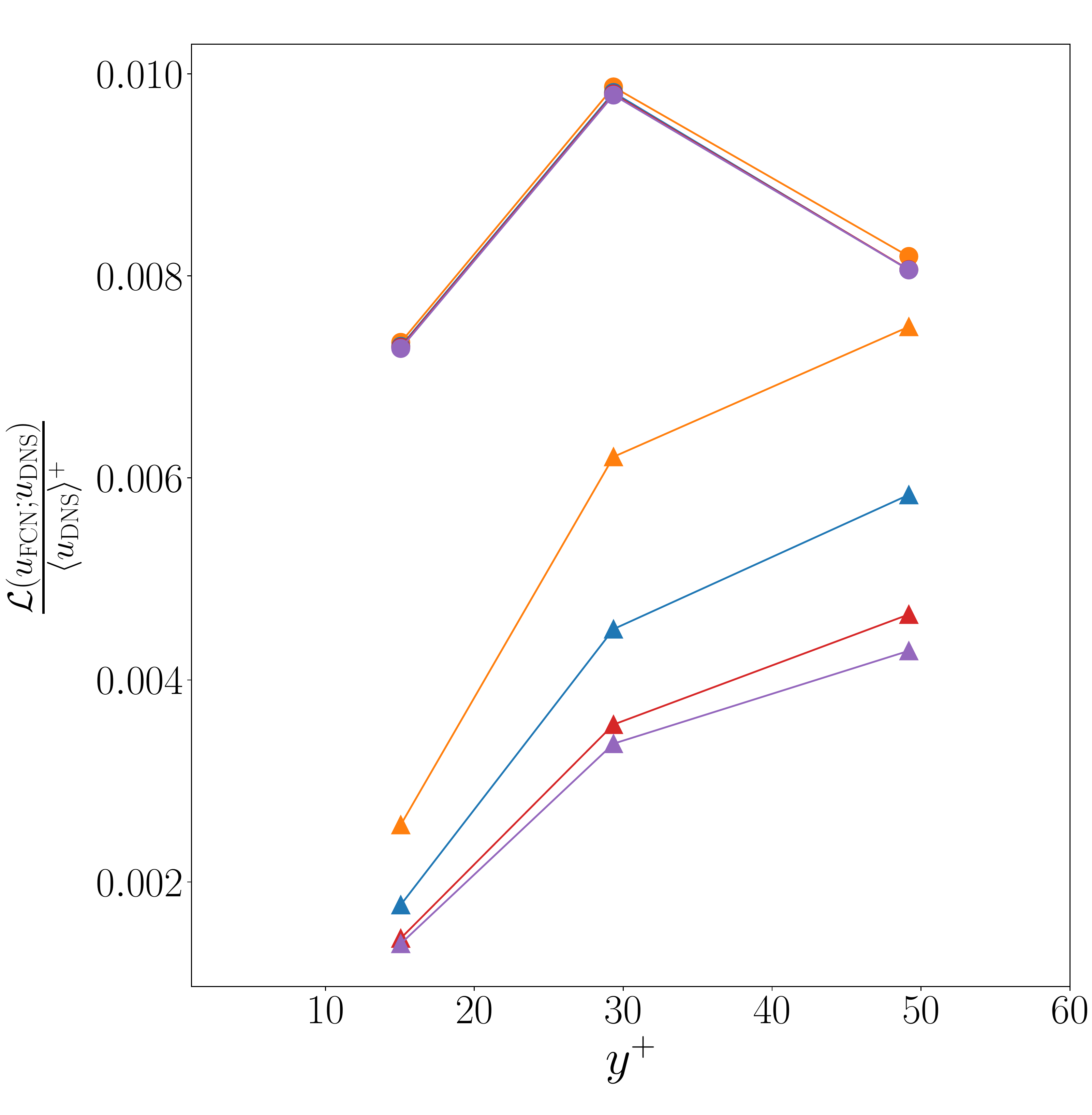}
\includegraphics[width=.328\textwidth]{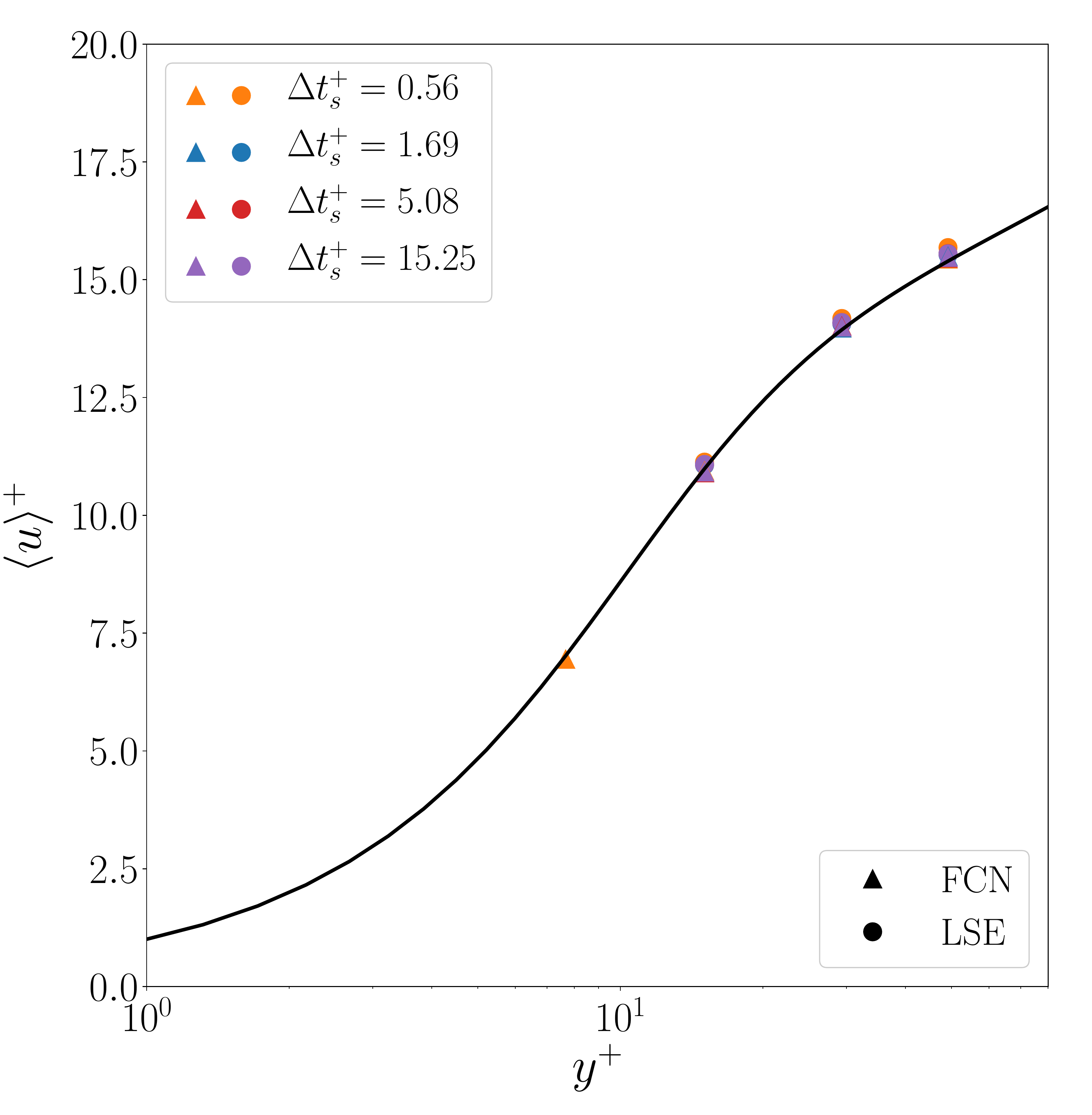}
\includegraphics[width=.328\textwidth]{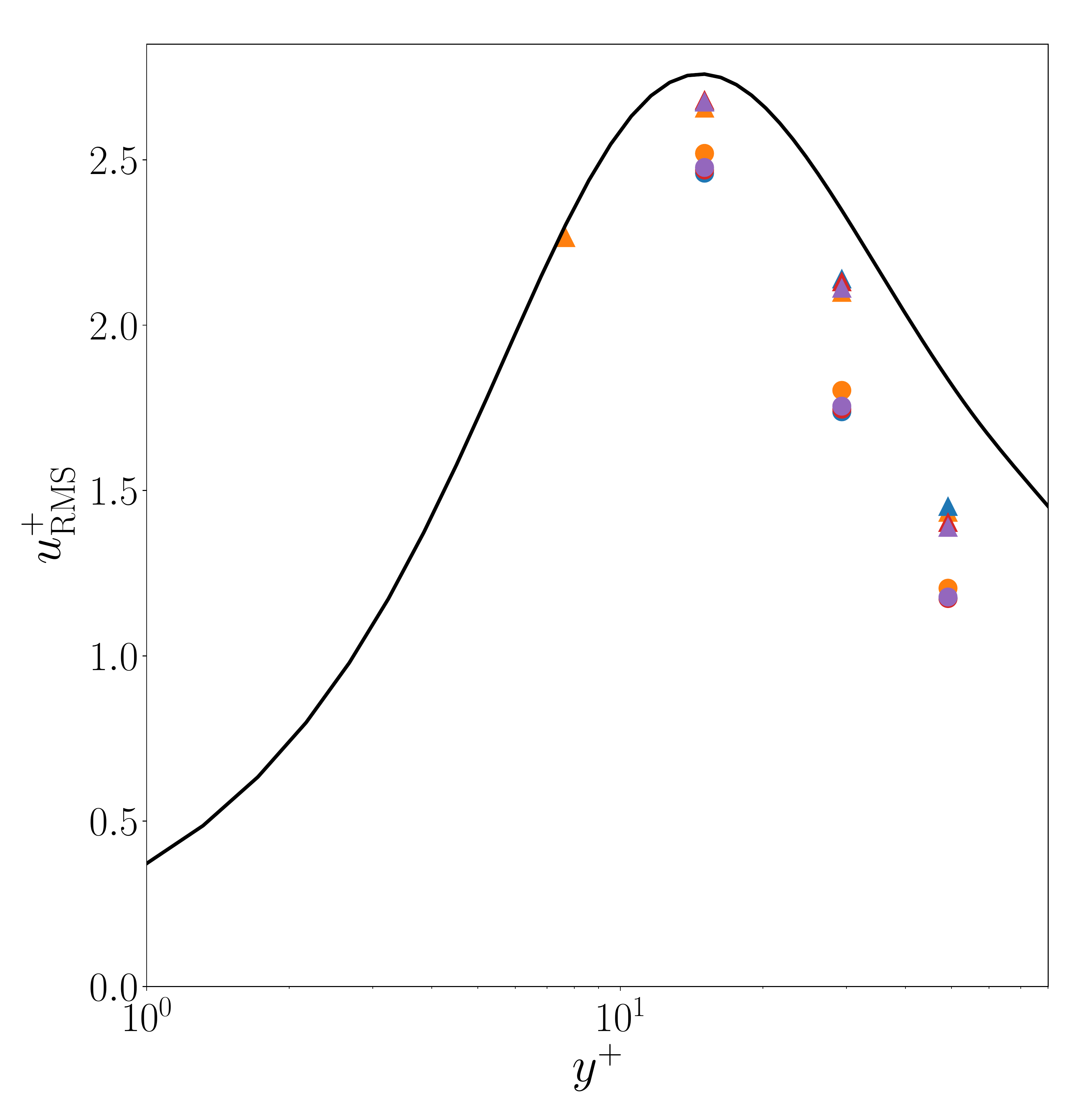}
\end{center}
\caption{\label{fig:stats} (Left) Validation loss  in the instantaneous fields predicted by the FCNs and the LSE with different time separations between samples, scaled with the mean velocity. (Middle) Comparison of the mean streamwise velocity and (right) the streamwise velocity fluctuations, between the DNS (\full) and the predicted flows with different $\Delta t^{+}_{s}$.} 
\end{figure}

Training on a dataset with less correlated samples leads to a higher training loss, because it is more difficult to find a combination of the parameters that fits all the training data. However, such a perfect fit is not always desirable, since it would most likely lead to a poor generalization performance (\textit{i.e.}\ the network would exhibit overfitting). This is clearly visible in Figure~\ref{fig:tstep}, when the samples are too correlated ({\it i.e.}\ for $\Delta t^{+}_{s}=0.56$ and 1.69). In those cases, both at $y^+=15$ (left) and $y^+=50$ (right), the validation loss is higher and increases as we iterate over the dataset, instead of being monotonically decreasing. Note that higher values of the time step between samples lead to smaller generalization gaps between the training and the validation losses. The training loss represents a lower bound for the error in the predictions of validation and test data. When $\Delta t^+_{s} = 15.25$ is used, the generalization gap is very small and for this reason it is unlikely to further improve the performance with the current architecture. Thus, the capacity of the network (\textit{i.e.}\ the number of trainable parameters in the network) becomes the limiting factor for the current predictions. Note that, to collect the same number of sample with increasing $\Delta t^+_{s}$, the overall simulation time, during which the sampling is performed, must increase as well.
The model-averaged validation loss at the end of the training, scaled with the mean velocity at the considered wall-normal location is shown in Figure~\ref{fig:stats} (left). 

After evaluating the accuracy of the instantaneous predictions, we assess the effect of the time step between samples on the ability of the network to reproduce the turbulence statistics.
The model-averaged mean and fluctuations for each $\Delta t^+_{s}$ are shown in Figure~\ref{fig:stats}~(middle and right). The mean velocity profile is accurately predicted for all $\Delta t^+_{s}$, with a consistent performance at all the considered wall-normal locations. On the other hand, it is possible to observe a performance degradation in the prediction of the streamwise velocity fluctuations when moving away from the wall. Even if a higher $\Delta t^+_{s}$ allows a more faithful prediction of the instantaneous flow fields, as shown by the reduction of the validation loss in Figure~\ref{fig:stats}~(left), there is little influence of this parameter on the statistics prediction, as shown in Table~\ref{tbl:stats}. 
In the same table, the error in the reconstruction of the LSE is reported. Also for this method we compared operators with different time step between samples and we found negligible differences both in the instantaneous (Fig.~\ref{fig:stats}, left) and the statistical accuracy (Fig.~\ref{fig:stats}, middle and right). For the LSE, the quality of the reconstructions improves with less correlated data until we obtain a converged operator. In our case, the dataset with $\Delta t^+_{s} = 0.56$ is already sufficient for the operator to reach convergence, hence in Table~\ref{tbl:stats} we show only the error of the LSE that reconstructs better the velocity-fluctuations profile. No standard deviation is available since the computation of the operator through the solution of the equation~\ref{eq:lse2} is a completely deterministic procedure.
As we observed qualitatively in Figure~\ref{fig:field_comp}, since the LSE reconstruction is limited to the linear part of the scale interactions, there is a significant underestimation of the fluctuations, especially away from the wall. The capability of the neural network to capture non-linear interactions helps to curb the loss of accuracy when the height of the predicted plane is increased. 

\begin{table}[h]
    \caption{\label{tbl:stats} Relative prediction error of the LSE and the neural networks, trained with different time steps between samples.} 
\begin{center}
\lineup
\begin{tabular}{l *{3}{c}}
\br
$E_{\langle u \rangle^+}\ [\%]$ & $y^+=15$ & $y^+=30$ & $y^+=50$  \cr 
\mr
LSE ($\Delta t^+_{s} = 0.56$) & $1.55$            & $1.33$   & $1.37$  \cr
FCN ($\Delta t^+_{s} = 0.56$)  & $0.81~(\pm 0.46)$ & $\00.52~(\pm 0.30)$ & $\00.48~(\pm 0.28)$  \cr
FCN ($\Delta t^+_{s} = 1.69$)  & $0.64~(\pm 0.35)$ & $\00.55~(\pm 0.32)$ & $\00.46~(\pm 0.37)$  \cr
FCN ($\Delta t^+_{s} = 5.08$)  & $0.68~(\pm 0.35)$ & $\00.59~(\pm 0.38)$ & $\00.54~(\pm 0.30)$  \cr
FCN ($\Delta t^+_{s} = 15.25$) & $0.66~(\pm 0.70)$ & $\00.81~(\pm 0.47)$ & $\00.31~(\pm 0.23)$  \cr
\br
\end{tabular}
\begin{tabular}{l *{3}{c}}
\br
$E_{u_\mathrm{RMS}^+}\ [\%]$ & $y^+=15$ & $y^+=30$ & $y^+=50$  \cr
\mr
LSE ($\Delta t^+_{s} = 0.56$) & $9.14$   & $24.5\0$ & $35.9$   \cr
FCN ($\Delta t^+_{s} = 0.56$)  & $2.87~(\pm 1.00)$   & $11.30~(\pm 0.80)$ & $23.15~(\pm 0.58)$   \cr
FCN ($\Delta t^+_{s} = 1.69$)  & $2.09~(\pm 0.50)$   & $\09.50~(\pm 1.26)$ & $22.10~(\pm 0.83)$   \cr
FCN ($\Delta t^+_{s} = 5.08$)  & $1.92~(\pm 0.98)$   & $\09.82~(\pm 0.75)$ & $24.66~(\pm 1.96)$   \cr
FCN ($\Delta t^+_{s} = 15.25$) & $2.11~(\pm 0.90)$   & $10.65~(\pm 2.11)$ & $25.57~(\pm 1.92)$   \cr
\br
\end{tabular}
\end{center}
\end{table}

\subsection{Variations on network architecture and training}\label{s:var}
\begin{table}[h]
    \caption{\label{tbl:multitask} Prediction error of the multiple-output neural networks, trained with different pairs of output heights, considering $\Delta t^+_{s} = 15.25$. The single-output (SO) networks trained with the same $\Delta t^+_{s}$ are included as reference.} 
\begin{center}
\lineup
\begin{tabular}{l *{3}{c}}
\br
$E_{\langle u \rangle^+}\ [\%]$ & $y^+=15$ & $y^+=30$ & $y^+=50$  \cr 
\mr
FCN SO               & $0.66~(\pm 0.70)$ & $\00.81~(\pm 0.47)$ & $\00.31~(\pm 0.23)$  \cr
FCN, $y^+=(15, 30)$  & $1.13~(\pm 0.67)$ & $\00.52~(\pm 0.53)$ & --                   \cr
FCN, $y^+=(15, 50)$  & $1.03~(\pm 0.25)$ & --                  & $\00.21~(\pm 0.22)$  \cr

\br
\end{tabular}
\begin{tabular}{l *{3}{c}}
\br
$E_{u_\mathrm{RMS}^+}\ [\%]$ & $y^+=15$ & $y^+=30$ & $y^+=50$  \cr
\mr
FCN SO               & $2.11~(\pm 0.90)$   & $10.65~(\pm 2.11)$ & $25.57~(\pm 1.92)$   \cr
FCN, $y^+=(15, 30)$  & $2.79~(\pm 0.80)$   & $10.54~(\pm 1.22)$ & --                   \cr
FCN, $y^+=(15, 50)$  & $1.79~(\pm 1.05)$   & --                 & $24.46~(\pm 1.68)$   \cr
\br
\end{tabular}
\end{center}
\end{table}

The multiple-output architecture discussed in Subsection~\ref{ss:nn} was tested for two pairs of target planes, namely $y^+=(15, 30)$ and $y^+=(15, 50)$. The statistics obtained from these networks are reported in Table~\ref{tbl:multitask}, and the results differ very little from the ones obtained with single-output networks. This finding supports the hypothesis that the distribution of the weights in the first layers of the network is very similar for all the different output heights. On the other hand, the parallel disposition of the branches does not provide a significant improvement over the previous results. Sharing the first layers only provides a small advantage from the computational performance perspective, since training a network with two outputs is about $12\%$ faster than training two separate single-output networks. However, if a reduction of the training time is sought after, a much greater improvement can be obtained by using transfer learning, discussed in Subsection~\ref{ss:nn}.
In our work, this technique has been applied to the training of a single-output network to predict the velocity field at $y^+=50$. A much faster learning rate decay has to be set, otherwise the optimization algorithm tends to move far away from the weight distribution suggested by the initialization, resulting in a rapid increase of the validation loss during training. The statistics results are compared with a network that shares almost all the configuration settings, except that the trainable parameters are randomly initialized and that all the network layers are updated through backpropagation. The results are summarized in Table~\ref{tbl:transfer}. With a training time that is more than 4 times lower, we are able to achieve a validation loss value that is as low as in the fully trainable case, predict planes that are qualitatively similar and obtain statistics with comparable accuracy, although the $E_{u_\mathrm{RMS}^+}$ is slightly higher. Provided the availability of training data for different planes, transfer learning can be used to obtain predictions at different planes in a very efficient way.

\begin{table}[h]
    \caption{\label{tbl:transfer} FCN validation loss and prediction errors for the velocity field at $y^+=50$, with random initialization (\textit{Fully trainable}) or with the model for $y^+=15$ used as initialization (\textit{Transfer learning}).} 

\begin{center}
\lineup
\begin{tabular}{l *{4}{c}}
\br                              
 & M.S.E. $[\times 10^{-3}]$ & $E_{\langle u \rangle^+}\ [\%]$ & $E_{u_\mathrm{RMS}^+}\ [\%]$ & Relative Time $[\%]$ \cr 
\mr
Fully trainable   & $3.04$ & $0.54~(\pm 0.30)$  & $24.66~(\pm 1.96)$ & $100$  \cr
Transfer learning & $3.17$ & $0.5$              & $30.2$    & $\023$ \cr 
\br
\end{tabular}
\end{center}
\end{table}

\section{Conclusions}\label{s:concl}
In this work, we assessed the possibility to predict a velocity  field from shear-stress measurements at the wall, using a fully-convolutional neural network (FCN). Both single- and multiple-output networks are able to provide predictions in good agreement with the flow fields computed by the DNS. The comparison of the statistics shows that the mean flow can be accurately computed from the predicted fields at all tested wall-normal locations. The streamwise velocity fluctuation profile is correctly reproduced, including the near-wall peak, however the $E_{u_\mathrm{RMS}^+}$ increases as we move away from the wall. Nevertheless, the capability of the FCN to predict the nonlinear scale interactions inherent to turbulence provides a clear advantage over linear methods, as shown in the comparison with LSE, which always exhibits higher reconstruction error. 

The quality of the instantaneous predictions improves if less correlated samples are used to train the FCNs, provided that the network capacity is sufficient to generalize over the training dataset. Also, the statistical accuracy does not show any dependency on the chosen $\Delta t^+_{s}$, which implies that for all applications requiring only the prediction of the statistical quantities, it is possible to reduce the computational cost of the generation of training data using a small time step between samples. Note however that the networks are not explicitly optimized to reproduce the statistics of the original simulation; to this end, the loss function should be modified to include new terms that account for such statistical quantities, thus minimizing the error in their prediction. 

The obtained results are promising, even though they are still preliminary and a more thorough investigation is required to completely uncover the capabilities of neural networks in this prediction task. To this end, the performance of the networks has to be assessed also at higher distances from the wall, where a larger separation between input samples and/or higher network capacity may be required for an accurate prediction.  

A current limitation of our model is that the networks have been trained using samples taken from a numerical simulation, where parameters like the mass flux, the viscosity and $Re_{\tau}$ can be accurately set and monitored. In order to deploy the trained model in a realistic situation, it will be necessary to test the robustness of the trained models to uncertainties of such parameters~\cite{REZAEIRAVESH201857}. 
Another limitation of our model is the lack of periodicity of the predicted fields, which will be addressed in future work. Currently, it prevents the direct utilization of the fields in the context of a numerical simulation, such as the DNS used to generate the training data.

The achieved results open different application possibilities for the trained models: experiments would benefit from the possibility to obtain the flow field predictions from quantities that can be easily measured, such as the wall-shear stress; 
for flow-control applications, the availability of a reliable and efficient representation of the flow field is fundamental to design an efficient controller.

\ack
This study was carried out in the context of the $4^{th}$ Madrid Turbulence Workshop, funded in part by the Coturb program of the European Research Council. The authors also acknowledge the funding provided by the Swedish e-Science Research Centre (SeRC) and the Knut and Alice Wallenberg (KAW) Foundation. Part of the analysis was performed on resources provided by the Swedish National Infrastructure for Computing (SNIC) at PDC. 
The authors would also like to acknowledge the help of Eiichi Sasaki during the review of the proceeding.
The code to generate Figure~\ref{fig:net} is adapted from \url{https://github.com/HarisIqbal88/PlotNeuralNet}.

\section*{References}
\bibliographystyle{iopart-num}
\bibliography{iopart-num}

\providecommand{\newblock}{}
\begin{thebibliography}{10}
\expandafter\ifx\csname url\endcsname\relax
  \def\url#1{{\tt #1}}\fi
\expandafter\ifx\csname urlprefix\endcsname\relax\def\urlprefix{URL }\fi
\providecommand{\eprint}[2][]{\url{#2}}

\bibitem{brunton2019}
Brunton S~L, Noack B~R and Koumoutsakos P 2020 {\em Annu. Rev. Fluid Mech.\/}
  {\bf 52}

\bibitem{ling_et_al}
Ling J, Kurzawski A and Templeton J 2016 {\em J. Fluid Mech.\/} {\bf 807}
  155--166

\bibitem{Rabault}
Rabault J and Kuhnle A 2019 {\em Phys. Fluids\/} {\bf 31} 094105

\bibitem{jimenez_ml}
Jim\'enez J 2018 {\em J. Fluid Mech.\/} {\bf 854, R1} 1--11

\bibitem{srinivasan}
Srinivasan P~A, Guastoni L, Azizpour H, Schlatter P and Vinuesa R 2019 {\em
  Phys. Rev. Fluids\/} {\bf 4} 054603

\bibitem{guastoni_tsfp}
Guastoni L, Srinivasan P~A, Azizpour H, Schlatter P and Vinuesa R 2019 {\em
  Proc. of the 11th Int. Symp. on Turbulence and Shear Flow Phenomena (TSFP11),
  Southampton, UK, July 30 - August 2.\/}

\bibitem{rumelhart1985learning}
Rumelhart D~E, Hinton G~E and Williams R~J 1985 Learning internal
  representations by error propagation Tech. rep. California Univ San Diego La
  Jolla Inst for Cognitive Science

\bibitem{hochreiter1997long}
Hochreiter S and Schmidhuber J 1997 {\em Neural comput.\/} {\bf 9} 1735--1780

\bibitem{moehlis_et_al}
Moehlis J, Faisst H and Eckhardt B 2004 {\em New J. Phys.\/} {\bf 6} 56

\bibitem{agostini_leschziner_2016}
Agostini L and Leschziner M 2016 {\em Phys. Fluids\/} {\bf 28} 015107

\bibitem{baars_et_al_2017}
Baars W~J, Hutchins N and Marusic I 2017 {\em Phil. Trans. R. Soc. A\/} {\bf
  375} 20160077

\bibitem{dogan_et_al_2019}
Dogan E, {\"O}rl{\"u} R, Gatti D, Vinuesa R and Schlatter P 2019 {\em Fluid
  Dyn. Res.\/} {\bf 51} 011408

\bibitem{dullerud}
Dullerud G~E and Paganini F 2000 {\em A Course in Robust Control Theory\/}
  (Springer)

\bibitem{illingworth_et_al_2018}
Illingworth S~J, Monty J~P and Marusic I 2018 {\em J. Fluid Mech.\/} {\bf 842}
  146--162

\bibitem{mckeon_sharma_2010}
McKeon B~J and Sharma A~S 2010 {\em J. Fluid Mech.\/} {\bf 658} 336--382

\bibitem{suzuki_hasegawa_2017}
Suzuki T and Hasegawa Y 2017 {\em J. of Fluid Mech.\/} {\bf 830} 760–796

\bibitem{encinar}
Encinar M~P, Lozano-Dur\'an A and Jim\'enez J 2018 {\em Center for Turbulence
  Research. Proc. of the Summer Program 2018\/} pp 217--226

\bibitem{sasaki_vinuesa_cavalieri_schlatter_henningson_2019}
Sasaki K, Vinuesa R, Cavalieri A~V~G, Schlatter P and Henningson D~S 2019 {\em
  J. Fluid Mech.\/} {\bf 864} 708–745

\bibitem{milano_koumoutsakos}
Milano M and Koumoutsakos P 2002 {\em J. Comput. Phys.\/} {\bf 182} 1--26

\bibitem{guemes}
G\"uemes A, Discetti S and Ianiro A 2019 {\em Phys. Fluids, Submitted\/}

\bibitem{Lumley}
Lumley J~L 1967 {\em Atmospheric turbulence and wave propagation, A. M. Yaglom
  and V. I. Tatarski (eds). Nauka, Moscow\/}  166--178

\bibitem{boree}
Bor{\'e}e J 2003 {\em Exp. Fluids\/} {\bf 35} 188--192

\bibitem{lozano_2012}
Lozano-Dur{\'a}n A, Flores O and Jim{\'e}nez J 2012 {\em J. Fluid Mech.\/} {\bf
  694} 100--130

\bibitem{chevalier}
Chevalier M, Schlatter P, Lundbladh A and Henningson D 2007 A pseudospectral
  solver for incompressible boundary layer flows Tech. rep. TRITA-MEK 2007:07.
  KTH Mechanics, Stockholm, Sweden

\bibitem{Lecun98gradient-basedlearning}
Lecun Y, Bottou L, Bengio Y and Haffner P 1998 {\em Proc. of the IEEE\/} pp
  2278--2324

\bibitem{Goodfellow:2016:DL:3086952}
Goodfellow I, Bengio Y and Courville A 2016 {\em Deep Learning\/} (The MIT
  Press)

\bibitem{dumoulin2016guide}
Dumoulin V and Visin F 2016 {\em arXiv preprint arXiv:1603.07285\/}

\bibitem{long2015fully}
Long J, Shelhamer E and Darrell T 2015 {\em Proc. of the IEEE Conf. on computer
  vision and pattern recognition\/} pp 3431--3440

\bibitem{oord2016wavenet}
Oord A~v~d, Dieleman S, Zen H, Simonyan K, Vinyals O, Graves A, Kalchbrenner N,
  Senior A and Kavukcuoglu K 2016
  \urlprefix\url{http://arxiv.org/abs/1609.03499}

\bibitem{Neil:2016:PLA:3157382.3157532}
Neil D, Pfeiffer M and Liu S~C 2016 {\em Proc. of the 30th Int. Conf. on Neural
  Information Processing Systems\/} NIPS'16 (USA: Curran Associates Inc.) pp
  3889--3897

\bibitem{kingmaba}
Kingma D~P and Ba J 2015 {\em 3rd Int. Conf. on Learning Representations,
  {ICLR} 2015, San Diego, CA, USA, May 7-9, 2015, Conference Track
  Proceedings\/}

\bibitem{ioffe}
Ioffe S and Szegedy C 2015 {\em Proc. of the 32nd Int. Conf. on Machine
  Learning - Volume 37\/} ICML'15 pp 448--456

\bibitem{nair2010rectified}
Nair V and Hinton G~E 2010 {\em Proc. of the 27th Int. Conf. on Machine
  Learning (ICML-10)\/} pp 807--814

\bibitem{azizpour2015factors}
Azizpour H, Razavian A~S, Sullivan J, Maki A and Carlsson S 2015 {\em IEEE
  transactions on pattern analysis and machine intelligence\/} {\bf 38}
  1790--1802

\bibitem{zhang2018unreasonable}
Zhang R, Isola P, Efros A~A, Shechtman E and Wang O 2018 {\em Proc. of the IEEE
  Conf. on Computer Vision and Pattern Recognition\/} pp 586--595

\bibitem{adrian88}
Adrian R 1988 {\em J. Fluid Mech.\/} {\bf 190} 531--559 ISSN 0022-1120

\bibitem{REZAEIRAVESH201857}
Rezaeiravesh S, Vinuesa R, Liefvendahl M and Schlatter P 2018 {\em Europ. J. of
  Mech. - B/Fluids\/} {\bf 72} 57 -- 73

\end{thebibliography}

\end{document}